\documentclass[12pt]{article}
\usepackage{a4wide}
\usepackage{latexsym}
\usepackage{cite}

\usepackage{pslatex}
\usepackage[latin1]{inputenc}
\usepackage[T1]{fontenc}

\newcommand{\bq}{\begin{eqnarray}}
\newcommand{\eq}{\end{eqnarray}}
\renewcommand{\l}{\langle}
\renewcommand{\r}{\rangle} 
\newcommand{\eps}{\varepsilon}

\usepackage{amsmath,amssymb}
\newcommand{\braket}[1]{\langle #1 \rangle}
\newcommand{\sbraket}[1]{\lbrack #1\rbrack}
\usepackage{cite}
\usepackage{mcite}
\numberwithin{equation}{section}

\def\bra#1{\mathinner{\langle{#1}|}}
\def\ket#1{\mathinner{|{#1}\rangle}}
\makeatletter
\newcommand{\fmslash}[2][0mu]{%
  \mathchoice
    {\fmsl@sh\displaystyle{#1}{#2}}%
    {\fmsl@sh\textstyle{#1}{#2}}%
    {\fmsl@sh\scriptstyle{#1}{#2}}%
    {\fmsl@sh\scriptscriptstyle{#1}{#2}}}
\newcommand{\fmsl@sh}[3]{%
  \m@th\ooalign{$\hfil#1\mkern#2/\hfil$\crcr$#1#3$}}
\makeatother

\begin{document}
\allowdisplaybreaks  
\thispagestyle{empty}

\begin{flushright}
  MZ-TH/07-02 \\
  PITHA-07/01\\
\end{flushright}

\vspace{1.5cm}

\begin{center}
  {\Large\bf On-shell recursion relations for all Born QCD amplitudes}\\
  \vspace{1cm}
  {\large Christian Schwinn}\\\hfill\\
      {\small \em
        Institut f\"ur Theoretische Physik E, RWTH Aachen,}\\
      {\small \em   D - 52056 Aachen, Germany}\\
  \vspace{1cm}
 \large{Stefan Weinzierl\\\hfill\\
      {\small \em Institut f{\"u}r Physik, Universit{\"a}t Mainz,}\\
      {\small \em D - 55099 Mainz, Germany}\\
  } 
\end{center}

\vspace{2cm}

\begin{abstract}\noindent
  {
We consider on-shell recursion relations for all Born QCD amplitudes.
This includes amplitudes with several pairs of quarks and massive quarks.
We give a detailed description on how to shift the external particles in spinor
space and clarify the allowed helicities of the shifted legs.
We proof that the corresponding meromorphic functions vanish at $z \rightarrow \infty$.
As an application we obtain compact expressions
for helicity amplitudes including a pair of massive quarks, one negative
helicity gluon and an arbitrary number of positive helicity gluons. 
   }
\end{abstract}

\vspace*{\fill}

\newpage

\section{Introduction}
\label{sect:intro}
In the past years, various new methods for efficient calculations in QCD have
been introduced, motivated by the relation of QCD amplitudes to twistor string
theory found in~\cite{Witten:2003nn}.  In particular these methods
include the
diagrammatic rules of
Cachazo, Svr\v{c}ek and Witten (CSW)~\cite{Cachazo:2004kj}, where tree level QCD
amplitudes are constructed from vertices that are off-shell continuations of
maximal helicity violating (MHV) amplitudes~\cite{Parke:1986gb}, and the
recursion relations of
Britto, Cachazo, Feng and Witten~(BCFW)~\cite{Britto:2004ap,Britto:2005fq} that construct
scattering amplitudes from on-shell amplitudes with external momenta shifted
into the complex plane.  The BCFW recursion relations have found numerous
applications in tree
level~\cite{Luo:2005rx,*Luo:2005my,Britto:2005dg,Badger:2005zh,Badger:2005jv,Forde:2005ue,Quigley:2005cu,Ferrario:2006np,Ozeren:2006ft,Dinsdale:2006sq,*Duhr:2006iq,*Draggiotis:2006er,deFlorian:2006ek,*deFlorian:2006vu}
and
one-loop~\cite{Bern:2005hs,Bern:2005ji,*Bern:2005cq,*Bern:2005hh,*Forde:2005hh,*Berger:2006ci,*Berger:2006vq,*Berger:2006sh}
calculations in QCD.  Extensions to QED~\cite{Ozeren:2005mp} and
gravity~\cite{Bedford:2005yy,Cachazo:2005ca,*Bjerrum-Bohr:2005jr,*Brandhuber:2007up,*Benincasa:2007qj} also have
been considered. The relation of the BCFW method to the usual Feynman diagrams
has been clarified in~\cite{Draggiotis:2005wq,*Vaman:2005dt} and it has been
used to give a proof for the CSW construction~\cite{Risager:2005vk}.
The main advantages of the BCFW and CSW constructions are in the
simplification of analytical calculations as compared to more
traditional off-shell recursive methods~\cite{Berends:1987me}.

Most of the literature related to these new methods restricts itself to the all-gluon amplitude.
For calculations of multi-parton scattering amplitudes relevant for
phenomenological applications at upcoming colliders such as the Large Hadron
Collider~(LHC) it is desirable to extent these methods towards the full particle content 
of the Standard Model.
In the CSW approach it has been possible
to include single external massive gauge bosons or Higgs
bosons~\cite{Dixon:2004za,*Badger:2004ty,*Bern:2004ba,*Badger:2006us}, while
the BCFW recursion relations have been successfully applied to derive
multigluon amplitudes involving a pair of massive
scalars~\cite{Badger:2005zh,Forde:2005ue}.  
As shown in~\cite{Schwinn:2006ca} some
helicity amplitudes for massive quarks can be obtained from these scalar
amplitudes by Ward-identities in super-symmetric-QCD.  
A compact
expression for amplitudes with a pair of massive scalars or quarks and an
arbitrary number of positive helicity gluons has been found
in~\cite{Rodrigo:2005eu,Ferrario:2006np} by a combination of off-shell recursive methods and
the BCFW relations.
 
On-shell recursion relations for amplitudes involving massive quarks
have been considered in~\cite{Badger:2005jv}, but the proof is
restricted to the case where the shifted particles are massless.
It should be noted that expressions for
shifts of massive momenta have been stated in~\cite{Badger:2005zh}.
However, Ozeren and Stirling~\cite{Ozeren:2006ft}
report that they were unable to construct all helicity combinations of the
$\bar{t} t \rightarrow g g g$ amplitude from on-shell recursion relations.
In addition, already the question of allowed helicities for the
shifts of massless quarks does not appear to be completely settled in the 
literature~\cite{Luo:2005rx,*Luo:2005my,Badger:2005jv,Quigley:2005cu,deFlorian:2006ek,*deFlorian:2006vu}.

The purpose of this paper is to clarify the situation for on-shell recursion relations for 
Born QCD amplitudes.
The particle content of QCD are gluons and quarks, where the latter may be massive or massless.
We derive expressions for shifts of spinors for all particles -- massless or not --
and investigate the allowed helicity combinations of the shifted
particles.  
Our findings can be summarised as follows: As in the massless case we have for each pair of marked
particles two possibilities to shift the spinors -- a holomorphic one and an anti-holomorphic one.
The two marked particles must not be quarks belonging to the same fermion line.
For each of the four possible helicity assignments of the two marked particles at least one shift
leads to a recursion relation.
The only exceptions to this rule are amplitudes involving solely massive quarks.
In this case two-particle shifts are not sufficient. 
However, amplitudes consisting only of massive quarks and sufficient many external legs may be computed
recursively from three-particle shifts.

We show that shifts of massive particles can lead to
 simpler recursion relations than those considered previously in the
 literature and use them to derive amplitudes with a pair of massive
 quarks, one negative helicity gluon and an arbitrary number of
 positive helicity gluons.  Using super-symmetric Ward
 identities we also obtain a more compact form for the
 corresponding amplitudes with a pair of massive scalars than known
 previously.

This paper is organised as follows: In section~\ref{sect:notation}
we introduce our notation 
together with a short review of the colour decomposition of QCD amplitudes 
and an introduction to the spinor helicity formalism.
Section~\ref{sect:recursion} explains in detail the recursion relation 
for Born QCD amplitudes.
This is the main result of this paper.
The proof of the recursion relation is given in section~\ref{sect:proof}.
In section~\ref{sect:applications} we discuss applications of the recursion relation and provide examples.
Section~\ref{sect:summary} contains our conclusions.
In appendix~\ref{appendix:spinors} we collected information on the construction of massless spinors out of light-like four-vectors.
Appendix~\ref{sect:exceptionalcases} contains the discussion of a few exceptional cases, which are needed for the proof
of the recursion relation in section~\ref{sect:proof}.

\section{Notation and conventions}
\label{sect:notation}

In this section we briefly review the colour decomposition of QCD amplitudes
and the spinor helicity formalism.

\subsection{Colour decomposition}

Amplitudes in QCD may be decomposed into group-theoretical factors (carrying the colour structures)
multiplied by kinematic functions called partial amplitudes
\cite{Cvitanovic:1980bu,*Berends:1987cv,*Mangano:1987xk,*Kosower:1987ic,*Bern:1990ux}. 
These partial amplitudes do not contain any colour information and are gauge-invariant objects. 
Although no arguments in this paper rely on colour decomposition, the examples we present are based
on partial amplitudes.
By convention we consider all particles as out-going.

In the pure gluonic case tree level amplitudes with $n$ external gluons may be written in 
the form
\bq
\label{colour_decomp_pure_gluon}
{\cal A}_{n}(1,2,...,n) & = & \left(\frac{g}{\sqrt{2}}\right)^{n-2} \sum\limits_{\sigma \in S_{n}/Z_{n}} 
 \delta_{i_{\sigma_1} j_{\sigma_2}} \delta_{i_{\sigma_2} j_{\sigma_3}} 
 ... \delta_{i_{\sigma_n} j_{\sigma_1}}  
 A_{n}\left( \sigma_1, ..., \sigma_n \right),
\eq
where the sum is over all non-cyclic permutations of the external gluon legs.
The quantities $A_n(\sigma_1,...,\sigma_n)$, called the partial amplitudes, contain the 
kinematic information.
They are colour-ordered, e.g. only diagrams with a particular cyclic ordering of the gluons contribute.
The choice of the basis for the colour structures is not unique, and several proposals
for bases can be found in the literature \cite{DelDuca:1999rs,Maltoni:2002mq}.
Here we use the ``colour-flow decomposition'' \cite{Maltoni:2002mq,Weinzierl:2005dd}.
As a further example we give the
the colour decomposition for a tree amplitude with a pair of quarks:
\bq
{\cal A}_{n+2}(q,1,2,...,n,\bar{q}) 
& = & \left(\frac{g}{\sqrt{2}}\right)^{n} 
 \sum\limits_{S_n} 
 \delta_{i_q j_{\sigma_1}} \delta_{i_{\sigma_1} j_{\sigma_2}} 
 ... \delta_{i_{\sigma_n} j_{\bar{q}}} 
A_{n+2}(q,\sigma_1,\sigma_2,...,\sigma_n,\bar{q}),
\eq
where the sum is over all permutations of the gluon legs. 
In squaring these amplitudes a colour projector
\bq
 \delta_{\bar{i} i} \delta_{j \bar{j}} - \frac{1}{N} \delta_{\bar{i} \bar{j} } \delta_{j i}
\eq
has to applied to each gluon.
While the colour structures of the examples quoted above are rather simple, the colour decomposition
can be become rather involved for amplitudes with many pairs of quarks.
A systematic algorithm for the colour decomposition and the diagrams contributing to a single colour
structure can be found in ref.\cite{Weinzierl:2005dd}.

While not strictly necessary, we consider in this paper only colour-ordered partial amplitudes.
These partial amplitudes are cyclic ordered within each colour cluster.
The cyclic order reduces significantly the number of possibilities of
dividing $n$ external particles into two set, such that particle $i$ belongs to one set, while particle
$j$ belongs to the other set.

\subsection{Spinors and polarisation vectors}

Let us consider two independent Weyl spinors $|q+\r$ and $\l q+|$. These two Weyl spinors define
a light-like four-vector
\bq
 q^\mu & = & \frac{1}{2} \l q+ | \gamma^\mu | q+ \r.
\eq
This four-vector can be used to associate to any not necessarily light-like four-vector $k$ a light-like
four-vector $k^\flat$:
\bq
\label{projection_null}
 k^\flat & = & k - \frac{k^2}{2 k \cdot q} q.
\eq
The four-vector $k^\flat$ satisfies $(k^\flat)^2=0$.
We can generalise this construction and associate to an arbitrary four-vector $K$
a four-vector $K^\flat_m$ defined through
\bq
\label{projection_massive}
 K^\flat_m & = & K - \frac{\left(K^2-m^2\right)}{2K \cdot q}q,
\eq
which satisfies 
\bq
 \left( K^\flat_m \right)^2 & = & m^2.
\eq
Therefore $K^\flat_m$ corresponds to the momentum of an on-shell particle with mass $m$.
It is worth noting that starting from $K$ and constructing directly a light-like four-vector $K^\flat$
through eq.~(\ref{projection_null}) is the same as first constructing $K^\flat_m$ by eq.~(\ref{projection_massive})
and then projecting $K^\flat_m$ onto
a light-like vector $(K^\flat_m)^\flat$:
\bq
 \left( K^\flat_m \right)^\flat & = & K^\flat.
\eq
The two Weyl spinors $|q+\r$ and $\l q+|$ are also used as reference spinors in the definition of the polarisations of the 
external particles. 
For massive fermions we take the spinors as~\cite{Schwinn:2005pi,Rodrigo:2005eu}
\bq
\label{eq:spinors} 
 u(-) = \frac{1}{\l p^\flat + | q - \r}  \left( p\!\!\!/ + m \right) | q - \r,
 & &
\bar{u}(+) = \frac{1}{\l q - | p^\flat + \r} \l q - | \left( p\!\!\!/ + m \right),
 \nonumber \\
 u(+) = \frac{1}{\l p^\flat - | q + \r} \left( p\!\!\!/ + m \right) | q + \r,
 & &
\bar{u}(-) = \frac{1}{\l q + | p^\flat - \r} \l q + | \left( p\!\!\!/ + m \right).
\eq
These expressions are similar to the ones introduced
in~\cite{Kleiss:1986qc}, the major difference is given by the fact, that the
denominators contain spinor products rather than ordinary square roots.
The spinor $u(p)$ corresponds to a particle with incoming momentum,
therefore it has the reversed helicity notation compared to the usual
conventions~\cite{Kleiss:1986qc,Ozeren:2006ft}. 
This notation will turn out to
simplify the discussion of the allowed helicity combinations of
shifted quark lines in section~\ref{sect:assembling} since the
same restrictions apply to outgoing quarks and incoming anti-quarks.
Furthermore, using the conventions~\eqref{eq:spinors} internal
propagators of quarks in the BCFW relation connect $+$ and $-$
labels, as in the gluon case.
We note for completeness that the spinors $v(\pm)$ and $\bar{v}(\pm)$ are given by
\bq
 v(\pm) = \frac{1}{\l p^\flat \mp | q \pm \r} \left( p\!\!\!/ - m \right) | q \pm \r,
&&
\bar{v}(\pm) = \frac{1}{\l q \mp | p^\flat \pm \r} \l q \mp | \left( p\!\!\!/ - m \right).
\eq
These spinors satisfy the Dirac equations
\bq 
\left( p\!\!\!/ - m \right) u(\lambda) = 0, & & 
\bar{u}(\lambda) \left( p\!\!\!/ - m \right) = 0,
\eq
the orthogonality relations
\bq
\bar{u}(\bar{\lambda}) u(-\lambda) & = & 2 m \delta_{\bar{\lambda}\lambda}, 
\eq
and the completeness relation
\bq
\sum\limits_{\lambda} u(-\lambda) \bar{u}(\lambda) & = & p\!\!\!/ + m.
\eq
We further have
\bq
 \bar{u}(\bar{\lambda}) \gamma^\mu u(-\lambda) & = & 2 p^\mu \delta_{\bar{\lambda} \lambda}.
\eq
In the massless limit the definition reduces to
\bq
 u(-) = | p+ \r,
 &&
 \bar{u}(+) = \l p+ |, 
 \nonumber \\
 u(+) = | p- \r,
 & &
 \bar{u}(-) = \l p- |,
\eq
and the spinors are independent of the reference spinors $|q+\r$ and $\l q+|$.
For the polarisation vectors of the gluons we take
\bq
\eps_{\mu}^{+} = \frac{\langle q-|\gamma_{\mu}|k-\rangle}{\sqrt{2} \langle q- | k + \rangle},
 & &
\eps_{\mu}^{-} = \frac{\langle q+|\gamma_{\mu}|k+\rangle}{\sqrt{2} \langle k + | q - \rangle}.
\eq
The dependence on the reference spinors which enters through the gluon polarisation vectors
will drop out in gauge invariant quantities.
In addition, as we have seen, the external spinors of massless fermions are explicitly independent of the reference
spinors. Therefore we find again that (gauge invariant) amplitudes will not depend on them.
However for massive fermions the reference spinors are related to the quantisation axis of the spin for 
this fermion, and the individual amplitudes with label $+$ or $-$ will therefore depend on the
reference spinors $|q+\r$ and $\l q+|$.
It is easy to relate helicity amplitudes of massive quarks
corresponding to one choice of reference spinors to another set of reference spinors.
If $|\tilde{q}+\r$ and $\l \tilde{q}+|$ is a second pair of reference spinors
we have the following transformation law 
\bq
\label{rotate_spin_1}
\left( \begin{array}{c}
 \bar{u}(+,\tilde{q}) \\
 \bar{u}(-,\tilde{q}) \\
 \end{array} \right)
 & = &
 \left( \begin{array}{cc}
 c_{11} & c_{12} \\
 c_{21} & c_{22} \\
 \end{array} \right)
\left( \begin{array}{c}
 \bar{u}(+,q) \\
 \bar{u}(-,q) \\
 \end{array} \right),
\eq
where
\bq
 c_{11} = \frac{\l \tilde{q}- | \fmslash p | q- \r}{\l \tilde{q} \tilde{p}^\flat \r [ p^\flat q ]},
 \;\;\;
 c_{12} = \frac{m \l \tilde{q} q \r}{\l \tilde{q} \tilde{p}^\flat \r \l p^\flat q \r},
 \;\;\;
 c_{21} = \frac{m [ \tilde{q} q ]}{[ \tilde{q} \tilde{p}^\flat] [ p^\flat q ]},
 \;\;\;
 c_{22} = \frac{\l \tilde{q}+ | \fmslash p | q+ \r}{[ \tilde{q} \tilde{p}^\flat] \l p^\flat q \r}.
\eq
Here, $\tilde{p}^\flat$ denotes the projection onto a light-like four-vector with respect to the reference vector
$\frac{1}{2} \l \tilde{q}+ | \gamma^\mu | \tilde{q} + \r$.
Similar, we have for an amplitude with an incoming massive quark
\bq
\label{rotate_spin_2}
\left( \begin{array}{c}
 u(+,\tilde{q}) \\
 u(-,\tilde{q}) \\
 \end{array} \right)
 & = &
 \left( \begin{array}{cc}
 c_{11} & -c_{12} \\
 -c_{21} & c_{22} \\
 \end{array} \right)
\left( \begin{array}{c}
 u(+,q) \\
 u(-,q) \\
 \end{array} \right).
\eq

\section{The recursion relation}
\label{sect:recursion}

In this section we state the on-shell recursion relation for Born QCD amplitudes.
Conventionally, an amplitude depends on a set of external momenta $\{p_1, p_2, ..., p_n\}$.
In a first step we replace each four-vector by two spinors and view a QCD amplitude as a function
of these spinors. In \ref{sect:amplitude_arguments} we show how to recover the original four-vectors
from the spinors.
The recursion relation shifts the spinors by massless spinors. Since we allow for massive external
particles, we have to associate to a pair of two external particles two pairs of massless spinors.
A convenient Lorentz-invariant solution is given in \ref{sect:choosing_spinors}.
With these spinors at hand we state the holomorphic shift and the anti-holomorphic shift 
in \ref{sect:holomorphic_shift} and
\ref{sect:antiholomorphic_shift}, respectively.
Finally, \ref{sect:assembling} assembles all ingredients and gives the recursion relation.
Here we also present a list of the allowed helicity combinations.
The proof of the recursion relation is deferred to section~\ref{sect:proof}.

\subsection{Arguments of the amplitudes}
\label{sect:amplitude_arguments}

To state the recursion relation it is best not to view a QCD amplitude
as a function of a set of four-momenta $\{p_1, p_2, ..., p_n\}$, but
to replace each four-vector $p_j$ by two spinors $u_j(-)$ and $\bar{u}_j(+)$.
It is sufficient to specify these two spinors, since the remaining spinors
$u_j(+)$ and $\bar{u}_j(-)$ can be obtained by raising and lowering 
dotted or undotted indices.
If we change for the moment from the bra-ket notation to the one with
dotted/undotted indices according to
\bq
|p+\rangle = p_A,         & & \langle p+| = p_{\dot{A}}, \\
|p-\rangle = p^{\dot{A}}, & & \langle p-| = p^A,
\eq
we have
\bq
 u(-) = p^\flat_A + \frac{m}{\l p^\flat + | q - \r} q^{\dot{B}},
 & &
 \bar{u}(+) = p^\flat_{\dot{A}} + \frac{m}{\l q - | p^\flat + \r} q^B.
\eq
$u(+)$ and $\bar{u}(-)$ are then given by
\bq
\label{reconstruct_ubarminus_uplus}
 \bar{u}(-) = p^{\flat A} + \frac{m}{\l q + | p^\flat - \r} q_{\dot{B}},
 & &
 u(+) = p^{\flat \dot{A}} + \frac{m}{\l p^\flat - | q + \r} q_B,
\eq
where
$p^{\flat A}$, $p^{\flat \dot{A}}$, $q_{\dot{B}}$ and $q_B$ are obtained as
\bq
 p^{\flat A} = \eps^{AB} p^\flat_B,
 \;\;\;
 p^{\flat \dot{A}} = \eps^{\dot{A}\dot{B}} p^\flat_{\dot{B}},
 \;\;\;
 q_{\dot{B}} = q^{\dot{A}} \eps_{\dot{A}\dot{B}},
 \;\;\;
 q_B = q^A \eps_{AB}.
\eq
The two-dimensional antisymmetric tensor is defined by
\bq
\varepsilon^{AB} = \varepsilon^{\dot{A}\dot{B}} = \varepsilon_{AB} = \varepsilon_{\dot{A}\dot{B}}
 =
\left(\begin{array}{cc}
 0 & 1\\
 -1 & 0 \\
\end{array} \right).
\eq
We see that $u(-)$ determines $\bar{u}(-)$, and that correspondingly
$\bar{u}(+)$ determines $u(+)$.
Given the spinors we obtain the four-vector $p^\mu$ as follows:
\bq
\label{reconstruct_momentum}
 p^\mu & = & \frac{1}{4} \; \mbox{Tr} \left( \gamma^\mu \sum\limits_{\lambda} u(-\lambda) \bar{u}(\lambda) \right).
\eq
Eq.~(\ref{reconstruct_momentum}) in combination with
eq.~(\ref{reconstruct_ubarminus_uplus}) allows the reconstruction of each 
four-vector $p^\mu_j$ from the two spinors $u_j(-)$ and $\bar{u}_j(+)$.

\subsection{Choosing the spinors}
\label{sect:choosing_spinors}

To derive the recursion relation we mark two particles $i$ and $j$, which need not be massless, with
four-momenta $p_i$ and $p_j$.
To these two four-momenta we associate two light-like four-momenta $l_i$ and $l_j$ as follows \cite{delAguila:2004nf,vanHameren:2005ed}:
If $p_i$ and $p_j$ are massless, $l_i$ and $l_j$ are given by
\bq
l_i = p_i, 
& &
l_j = p_j.
\eq
If $p_i$ is massless, but $p_j$ is massive one has
\bq
l_i = p_i, 
\;\;\;
l_j = -\alpha_i p_i + p_j,
\;\;\;
\alpha_i = \frac{p_j^2}{2p_i p_j}.
\eq
The inverse formula is given by
\bq
p_i = l_i, 
& &
p_j = \alpha_i l_i + l_j.
\eq
If both $p_i$ and $p_j$ are massive,
one has
\bq
l_i = \frac{1}{1-\alpha_i \alpha_j} \left( p_i - \alpha_j p_j \right), 
& &
l_j = \frac{1}{1-\alpha_i \alpha_j} \left( -\alpha_i p_i + p_j \right).
\eq
$\alpha_1$ and $\alpha_2$ are given by
\bq
\alpha_j = \frac{2p_ip_j-\mbox{sign}(2p_ip_j)\sqrt{\Delta}}{2p_j^2},
 & &
\alpha_i = \frac{2p_ip_j-\mbox{sign}(2p_ip_j)\sqrt{\Delta}}{2p_i^2}.
\eq
Here,
\bq
\Delta & = & \left( 2p_ip_j \right)^2 - 4p_i^2 p_j^2.
\eq
The signs are chosen in such away that the massless limit $p_i^2 \rightarrow 0$ (or $p_j^2 \rightarrow 0$)
are approached smoothly.
The inverse formula is given by
\bq
\label{decompmom}
p_i = l_i + \alpha_j l_j, 
& &
p_j = \alpha_i l_i + l_j.
\eq
Note that $l_1$, $l_2$ are real for $\Delta>0$. 
For $\Delta<0$, $l_1$ and $l_2$ acquire imaginary parts.
As a summary we can associate to any pair $(p_i,p_j)$ of four-vectors a pair of
light-like four-vectors $(l_i,l_j)$.
These light-like four-vectors define massless spinors
$|l_i+ \r$, $\l l_i+ |$, $|l_j+ \r$ and $\l l_j+ |$.
Explicit formulae for the construction of these spinors are given in appendix \ref{appendix:spinors}.

\subsection{The holomorphic shift}
\label{sect:holomorphic_shift}

In an amplitude we single out two particles (massive or not) for special treatment.
From the four-vectors $p_i$ and $p_j$ we first obtain the two light-like four-vectors $l_i$ and $l_j$
and the associated massless spinors $|l_i+ \r$, $\l l_i+ |$, $|l_j+ \r$ and $\l l_j+ |$.
We consider helicity amplitudes. 
If particle $i$ is a massive quark or anti-quark, we use 
$|l_j+ \r$ and $\l l_j+ |$ as reference spinors for particle $i$.
If particle $j$ is a massive quark or anti-quark, we use
$|l_i+ \r$ and  $\l l_i+ |$ as reference
spinors for particle $j$.
In this case it is an easy exercise to show that
\bq
 p_i^\flat = l_i,
 & & 
 p_j^\flat = l_j.
\eq
For massive particles the reference
momenta define the spin quantisation axis.
If particle $i$ or $j$ is a gluon, we leave the corresponding reference spinors unspecified.
Gauge invariant quantities do not depend on the choice of reference spinors for gluons.
In the rest of the paper we will often choose specific reference spinors. 
It should be understood that this choice only affects massive quarks or anti-quarks.
The spinors read in detail:
\bq 
\label{eq:holomorphic}
 u_i(-) = | l_i + \r + \frac{m_i}{[ l_i l_j ]}  | l_j - \r,
 & &
\bar{u}_i(+) = \l l_i + | + \frac{m_i}{\l l_j l_i \r} \l l_j - |,
 \nonumber \\
 u_j(-) = | l_j + \r + \frac{m_j}{[ l_j l_i ]}  | l_i - \r,
 & &
\bar{u}_j(+) = \l l_j + | + \frac{m_j}{\l l_i l_j \r} \l l_i - |.
\eq
For the holomorphic shift 
we shift $u_i(-)$ and $\bar{u}_j(+)$, while $u_j(-)$ and $\bar{u}_i(+)$ remain unchanged: 
\bq
\label{shift_holomorphic}
 {u_i}'(-) = u_i(-) - z | l_j + \r,
 & &
 {\bar{u}_i}'(+) = \bar{u}_i(+),
 \nonumber \\
 {u_j}'(-) = u_j(-),
 & &
 {\bar{u}_j}'(+) = \bar{u}_j(+) + z \l l_i + |.
\eq
If both particles are massless we have $l_i=p_i$ and $l_j=p_j$. Then the shift defined in eq.~(\ref{shift_holomorphic})
reduces to the well-known form
\bq
 |p_i' +\r = |p_i +\r - z | p_j + \r,
 & &
 \l p_i'+ | = \l p_i +|,
 \nonumber \\
 | p_j' + \r = | p_j + \r,
 & &
 \l p_j ' +| = \l p_j + | + z \l p_i + |.
\eq
The spinors ${u_i}'(-)$ and ${\bar{u}_i}'(+)$ correspond to an on-shell particle with mass $m_i$ and four-momentum
\bq
 {p_i'}^\mu & = & p_i^\mu - \frac{z}{2} \left\l l_i+ \left| \gamma^\mu \right| l_j+ \right\r.
\eq
The spinors ${u_j}'(-)$ and ${\bar{u}_j}'(+)$ correspond to an on-shell particle with mass $m_j$ and four-momentum
\bq
 {p_j'}^\mu & = & p_j^\mu + \frac{z}{2} \left\l l_i+ \left| \gamma^\mu \right| l_j+ \right\r.
\eq
It is worth to examine the requirement to use $|l_j+\r$ and $\l l_j+|$ as reference
spinors for particle $i$ in detail.
Assume that we have an arbitrary spin quantisation axis described by the reference spinors
$|q+\r$ and $\l q+|$.
As before we perform the shift
\bq
 {u_i}'(-) = u_i(-) - z | l_j + \r,
 & &
 {\bar{u}_i}'(+) = \bar{u}_i(+).
\eq
If we now consider the polarisation sum we find
\bq
 \sum\limits_{\lambda} u_i'(-\lambda) \bar{u}_i'(\lambda)
 & = &
 p\!\!\!/_i + m_i
 - z \left(
 \begin{array}{cc}
 \frac{m_i}{\l p_i^\flat q \r} \left( | q+ \r \l l_j - | - | l_j + \r \l q- | \right) & | l_j+ \r \l p_i^\flat+ |  \\
 | p_i^\flat- \r \l l_j - | & 0 \\
 \end{array}
 \right).\;\;\;\;\;\;
\eq
As this polarisation sum must have the form
\bq
 p\!\!\!/_i' + m_i,
\eq
we have to require that the entry in the upper left corner vanishes:
\bq
 | q+ \r \l l_j - | \; - \; | l_j + \r \l q- | & = & 0.
\eq
Therefore it follows that $|q+\r=\lambda |l_j+\r$. The requirement $\l q+|=\lambda' \l l_j+|$
follows from similar considerations related to the anti-holomorphic shift discussed in the next sub-section.
Because not all helicity combinations can be computed with the holomorphic shift,
we have to fix both reference spinors $|q+\r$ and $\l q+|$, 
and use the anti-holomorphic shift as well as the holomorphic shift to compute all helicity combinations.
This allows us to recover the helicity
amplitudes for arbitrary reference spinors from~\eqref{rotate_spin_1} and~\eqref{rotate_spin_2} .

Finally we remark that the spinors should not lead to spurious poles in $z$
in the analytically continued scattering amplitude $A(z)$.
This excludes for instance the choice
$u_i'(-)=(\fmslash p'_i+m)|q-\r/[ p_i^\flat q ]$ and 
$\bar u_i'(+)= \l q-| (\fmslash p'_i+m)/(\l q p_i^\flat\r - z \l q l_j\r) $ for
$| q+ \r \neq | l_j+ \r$.
\\
\\
Let $k$ be an intermediate particle where we would like to factorise the amplitude.
We denote by $K$ the off-shell four-momentum flowing through this propagator in the
unshifted amplitude.
We define the polarisations with respect to the reference spinors $|l_j + \r$ and $\l l_i + |$:
\bq
\label{eq:inner-spinor}
 {u_K}'(-) = \frac{1}{\l K^\flat + | l_i - \r} \left( K\!\!\!/' + m_k \right) \left| l_i - \right\r,
 & &
 {\bar{u}_K}'(+) = \frac{1}{\l l_j- | K^\flat + \r} \left\l l_j- \right| \left( K\!\!\!/' + m_k \right),
\eq
where
\bq
 {K'}^\mu = K^\mu - \frac{z}{2} \l l_i+ | \gamma^\mu | l_j + \r,
 & &
 {K^\flat}^\mu = K^\mu - \frac{1}{2} \frac{K^2}{\l l_i+| K | l_j+ \r} \l l_i+ | \gamma^\mu | l_j + \r.
\eq
$K^\flat$ is a light-like four-vector. Note that $K^\flat = (K')^\flat$.
Furthermore $K'$ is on-shell ($(K')^2=m_k^2$) provided 
\bq
\label{z_holomorphic}
 z & = & \frac{K^2-m_k^2}{\l l_i+| K | l_j+ \r}.
\eq

\subsection{The anti-holomorphic shift}
\label{sect:antiholomorphic_shift}

For the anti-holomorphic shift we modify $\bar{u}_i(+)$ and $u_j(-)$:
\bq
\label{shift_antiholomorphic}
 {u_i}'(-) = u_i(-),
 & &
 {\bar{u}_i}'(+) = \bar{u}_i(+) - z \l l_j+ |,
 \nonumber \\
 {u_j}'(-) = u_j(-) + z | l_i + \r,
 & &
 {\bar{u}_j}'(+) = \bar{u}_j(+).
\eq
If both particles are massless the shift defined in eq.~(\ref{shift_antiholomorphic})
reduces to the form
\bq
 |p_i' +\r = |p_i +\r,
 & &
 \l p_i'+ | = \l p_i +| - z \l p_j +|,
 \nonumber \\
 | p_j' + \r = | p_j + \r + z | p_i + \r,
 & &
 \l p_j ' +| = \l p_j + |.
\eq
The spinors ${u_i}'(-)$ and ${\bar{u}_i}'(+)$ correspond to an on-shell particle with mass $m_i$ and four-momentum
\bq
 {p_i'}^\mu & = & p_i^\mu - \frac{z}{2} \left\l l_j+ \left| \gamma^\mu \right| l_i+ \right\r.
\eq
The spinors ${u_j}'(-)$ and ${\bar{u}_j}'(+)$ correspond to an on-shell particle with mass $m_j$ and four-momentum
\bq
 {p_j'}^\mu & = & p_j^\mu + \frac{z}{2} \left\l l_j+ \left| \gamma^\mu \right| l_i+ \right\r.
\eq
Again, let $k$ be an intermediate particle 
with off-shell four-momentum $K$.
We define the polarisations now with respect to the reference spinors $|l_i + \r$ and $\l l_j + |$:
\bq
 {u_K}'(-) = \frac{1}{\l K^\flat + | l_j - \r} \left( K\!\!\!/' + m_k \right) \left| l_j - \right\r,
 & &
 {\bar{u}_K}'(+) = \frac{1}{\l l_i- | K^\flat + \r} \left\l l_i- \right| \left( K\!\!\!/' + m_k \right),
\eq
where
\bq
 {K'}^\mu = K^\mu - \frac{z}{2} \l l_j+ | \gamma^\mu | l_i + \r,
 & &
 {K^\flat}^\mu = K^\mu - \frac{1}{2} \frac{K^2}{\l l_j+| K | l_i+ \r} \l l_j+ | \gamma^\mu | l_i + \r.
\eq
$K^\flat$ is a light-like four-vector
and we have $K^\flat = (K')^\flat$.
Furthermore $K'$ is on-shell ($(K')^2=m_k^2$) provided 
\bq
\label{z_antiholomorphic}
 z & = & \frac{K^2-m_k^2}{\l l_j+| K | l_i+ \r}.
\eq

\subsection{Assembling the ingredients: the recursion relation}
\label{sect:assembling}

We can now state the recursion relation. 
The starting point is the function
\bq
 A(z) & = & 
A_n\left( u_1(-), \bar{u}_1(+), \lambda_1, ..., u_i'(-), \bar{u}_i'(+), \lambda_i, ..., u_j'(-), \bar{u}_j'(+), \lambda_j, ... \right),
\eq
where the spinors of particles $i$ and $j$ have been shifted either with the holomorphic shift
or with the anti-holomorphic shift.
The amplitude we want to calculate is given by $A(0)$.
If the shifted amplitude $A(z)$ vanishes for $z\rightarrow \infty$ we obtain:
\bq
\label{on_shell_recursion}
\lefteqn{
A_n\left( u_1(-), \bar{u}_1(+), \lambda_1, ..., u_n(-), \bar{u}_n(+), \lambda_n\right)
 = 
 } & & \\
 & & 
\hspace*{20mm}
 \sum\limits_{partitions} \sum\limits_{\lambda=\pm}
  A_{L}\left( ..., u_i'(-), \bar{u}_i'(+), \lambda_i, ..., 
              i v_K'(-), i \bar{v}_K'(+), -\lambda
              \right)
 \nonumber \\
 & &
\hspace*{20mm}
  \times
  \frac{i}{K^2-m_k^2} 
  A_{R}\left( u_K'(-), \bar{u}_K'(+), \lambda,
              ..., u_j'(-), \bar{u}_j'(+), \lambda_j, ...
              \right),
 \nonumber 
\eq
where the sum is over all partitions such that particle $i$ is on the left and particle $j$ is on the right.
The momentum $K$ is given as the sum over all unshifted momenta of the original 
external particles, which are part of $A_L$.
The values of $z$ are given for the holomorphic shift by eq.~(\ref{z_holomorphic}) and for the anti-holomorphic shift 
by eq.~(\ref{z_antiholomorphic}).
\\
\\
The condition that $A(z)$ has to vanish at infinity can be summarised as follows:
\begin{itemize}
\item Particles $i$ and $j$ cannot belong to the same fermion line.
\item The holomorphic shift can be used for the helicity combinations
$(i^+,j^-)$, $(i^+,j^+)$ and $(i^-,j^-)$ with the following exceptions:
\begin{itemize}
\item The combinations $(q_i^+,g_j^+)$, $(\bar{q}_i^+,g_j^+)$, $(g_i^-,q_j^-)$ 
and $(g_i^-,\bar{q}_j^-)$ are not allowed.
\item If particle $i$ is a massive quark or anti-quark, the combinations
$(q_i^+,q_j'{}^+)$, $(q_i^+,\bar{q}_j'{}^+)$,  
$(\bar{q}_i^+,q_j'{}^+)$ and $(\bar{q}_i^+,\bar{q}_j'{}^+)$ are not allowed.  
\item If particle $j$ is a massive quark or anti-quark, the combinations
$(q_i^-,q_j'{}^-)$, $(q_i^-,\bar{q}_j'{}^-)$,  
$(\bar{q}_i^-,q_j'{}^-)$ and $(\bar{q}_i^-,\bar{q}_j'{}^-)$ are not allowed.  
\end{itemize}
\item The anti-holomorphic shift can be used for the helicity combinations
$(i^-,j^+)$, $(i^+,j^+)$ and $(i^-,j^-)$ with the following exceptions:
\begin{itemize}
\item The combinations $(g_i^+,q_j^+)$, $(g_i^+,\bar{q}_j^+)$, $(q_i^-,g_j^-)$ 
and $(\bar{q}_i^-,g_j^-)$ are not allowed.
\item If particle $j$ is a massive quark or anti-quark, the combinations
$(q_i^+,q_j'{}^+)$, $(q_i^+,\bar{q}_j'{}^+)$,  
$(\bar{q}_i^+,q_j'{}^+)$ and $(\bar{q}_i^+,\bar{q}_j'{}^+)$ are not allowed.  
\item If particle $i$ is a massive quark or anti-quark, the combinations
$(q_i^-,q_j'{}^-)$, $(q_i^-,\bar{q}_j'{}^-)$,  
$(\bar{q}_i^-,q_j'{}^-)$ and $(\bar{q}_i^-,\bar{q}_j'{}^-)$ are not allowed.  
\end{itemize}
\end{itemize}
In summary there is always at least one allowed shift, unless $i$ and $j$ belong to the same
fermion line or $i$ and $j$ are both massive quarks or anti-quarks.
As we are free to choose the particles $i$ and $j$, we can compute all Born helicity amplitudes 
in QCD with two-particle shifts via recursion relations, except the ones which involve only
massive quarks or anti-quarks.
The latter ones may be calculated recursively if one allows more general shifts, where more than
two particles are shifted. This follows directly from the proof of the recursion relation which
we present in section \ref{sect:proof}.
Amplitudes with only massive quarks or anti-quarks are discussed in detail 
in section \ref{sect:only_massive_quarks}.

\section{Proof of the recursion relation}
\label{sect:proof}

The standard proof of the BCFW recursion relation is based on Cauchy's theorem \cite{Britto:2005fq}.
The function $A(z)$ is a rational function of $z$, which has only simple poles in $z$.
Therefore, if $A(z)$ vanishes for $z \rightarrow \infty$, $A(z)$ is given 
by Cauchy's theorem as the sum over its residues. This is just the right hand side of the recursion relation.
The essential ingredient for the proof is the vanishing of $A(z)$ at $z \rightarrow \infty$.
This property we have to verify for the shifts stated in the previous section.

It is relatively easy to show this for the helicity combination $(i^+,j^-)$ for the holomorphic shift
and for the helicity combination $(i^-,j^+)$ for the anti-holomorphic shift. 
We do this in section~\ref{sect:diagrammatic_analysis}.

The helicity combinations $(i^+,j^+)$ and $(i^-,j^-)$ require a more sophisticated proof.
In section~\ref{supplementary_recursion} 
we first construct a representation of $A(z)$ with the help of a
supplementary recursion relation and deduct from this representation the large $z$-behaviour of $A(z)$.
For the proof we borrowed ideas 
from \cite{Badger:2005zh,Bern:2005hs,Bedford:2005yy,Risager:2005vk}.
The proof presented here does not rely on additional (unnecessary) assumptions like
the presence of two additional gluons with specific helicities.

\subsection{Diagrammatic analysis of the large $z$ behaviour}
\label{sect:diagrammatic_analysis}

We now investigate the behaviour of $A(z)$ for large $z$ by a diagrammatic analysis.
A gluon propagator behaves like $1/z$, whereas a quark propagator tends towards a constant.
The quark-gluon and the four-gluon vertices are independent of $z$, whereas the three-gluon vertex is 
proportional to $z$ for large $z$.
\begin{table}
\begin{center}
\begin{tabular}{|r|llllll|llllll|}
\hline
 & $g_i^+$ & $g_i^-$ & $\bar{Q}_i^+$ & $\bar{Q}_i^-$ & $Q_i^+$ & $Q_i^-$ & $g_j^+$ & $g_j^-$ & $\bar{Q}_j^+$ & $\bar{Q}_j^-$ & $Q_j^+$ & $Q_j^-$ \\  
 \hline
 & & & & & & & & & & & & \\  
holomorphic & $\frac{1}{z}$ & $z$ & $1$ & $z$ & $1$ & $z$ & $z$ & $\frac{1}{z}$ & $z$ & $1$ & $z$ & $1$ \\
 & & & & & & & & & & & & \\  
anti-holomorphic & $z$ & $\frac{1}{z}$ & $z$ & $1$ & $z$ & $1$ & $\frac{1}{z}$ & $z$ & $1$ & $z$ & $1$ & $z$\\
 & & & & & & & & & & & & \\  
\hline
\end{tabular}
\caption{\label{table1}
Behaviour of polarisation vectors and spinors in the large $z$ limit for the holomorphic and
anti-holomorphic shift.
}
\end{center}
\end{table}
The behaviour of the polarisation vectors and spinors are summarised in table~\ref{table1}.

As a first observation we note 
that a shift of two quarks belonging to the same fermion line is not allowed.
In all diagrams the $z$-dependence flows along this fermion line, which consists of
quark propagators and quark-gluon vertices. 
These tend towards a constant for large $z$.
The large $z$ behaviour of the external spinors tends towards a constant at the best.
Therefore, we conclude that independent of the helicities the function $A(z)$ does not vanish for
$z \rightarrow \infty$.

Let us now assume that the two shifted particles belong to different fermion lines, 
or that one or both particles are gluons.
Therefore we have at least one gluon propagator along the shifted line, except for the case where
the shifted line does not contain any propagators at all.
By a diagrammatic analysis one can easily show that the helicity combination $(i^+,j^-)$ 
behaves like $1/z$ for $z\rightarrow \infty$ for the holomorphic shift, 
independent of the nature of the particles
$i$ and $j$.
The reversed helicity assignment $(i^-,j^+)$ behaves like
$1/z$ for the anti-holomorphic shift.
To see this, let us consider as an example the holomorphic shift. 
Assume first that the flow of $z$-dependence in a particular diagram is given by a path made out 
entirely of gluons. The most dangerous contribution comes from a path, where all vertices are
three-gluon-vertices. For a path made of $n$ propagators we have $n+1$ vertices and the product of propagators
and vertices behaves therefore like $z$ for large $z$. 
This statement remains true for a path containing only one vertex and no propagators.
The polarisation vectors for the
helicity combination $(i^+,j^-)$ contribute a factor $1/z^2$, therefore the complete diagram
behaves like $1/z$ and vanishes therefore for $z \rightarrow \infty$.
If internally a gluon propagator is replaced by a quark propagator, we have to change at least
two three-gluon vertices into quark-gluon vertices. This improves the estimate by a factor $1/z$.
If an external gluon is replaced by a fermion, we have to change at least one three-gluon vertex into a
quark-gluon vertex. This does not modify the large $z$ behaviour.

\subsection{Supplementary recursion relation for the large $z$ behaviour}
\label{supplementary_recursion}

The cases $(i^+,j^+)$ and $(i^-,j^-)$ are more subtle.
As an example we consider the case $(i^+,j^+)$ with the holomorphic shift.
The other cases, $(i^+,j^+)$ with the anti-holomorphic shift and  
$(i^-,j^-)$ with the holomorphic as well as with the anti-holomorphic shift will be similar.

We are going to prove that in the case $(i^+,j^+)$ and for the holomorphic shift 
the function $A(z)$ vanishes as $z\rightarrow \infty$.
We prove this for the case where the spinors
$u_i(-)$, $\bar u_i(+)$, $u_j(-)$ and $\bar u_j(+)$ 
are defined with respect to the reference spinors
\bq
\label{choice_of_reference_spinors}
 | q_i + \r = | l_j + \r,
 & & 
 \l q_j + | = \l l_i + |.
\eq
Compared to section~\ref{sect:holomorphic_shift} we do not require any particular choice
for the reference spinors $\l q_i + |$ and $| q_j + \r$.
We can write these last two reference spinors as linear combinations of two basis spinors, and since
we are free to choose the normalisation of the reference spinors we can write them
without loss of generality as
\bq
 \l q_i + | = \l l_j + | + \lambda_i \l l_i + |,
 & &
 | q_j + \r = | l_i + \r + \lambda_j | l_j  \r,
\eq
where $\lambda_i$ and $\lambda_j$ are complex numbers.
A simple calculation shows that we then obtain with these reference spinors
\bq
 | p_i^\flat + \r = | l_i + \r - \lambda_i \frac{m_i^2}{2l_il_j} | l_j + \r,
 & & 
 \l p_i^\flat + | = \l l_i + |,
 \nonumber \\
 | p_j^\flat + \r = | l_j + \r,
 & &
 \l p_j^\flat + | = \l l_j + | - \lambda_j \frac{m_j^2}{2l_il_j} \l l_i + |.
\eq
The spinors $u_i(-)$, $\bar{u}_i(+)$, $u_j(-)$ and $\bar{u}_j(+)$ read then
\bq 
 u_i(-) = | p_i^\flat + \r + \frac{m_i}{[ l_i q_i ]}  | q_i - \r,
 & &
\bar{u}_i(+) = \l l_i + | + \frac{m_i}{\l l_j l_i \r} \l l_j - |,
 \nonumber \\
 u_j(-) = | l_j + \r + \frac{m_j}{[ l_j l_i ]}  | l_i - \r,
 & &
\bar{u}_j(+) = \l p_j^\flat + | + \frac{m_j}{\l q_j l_j \r} \l q_j - |.
\eq
The holomorphic shift is chosen as in eq.~(\ref{shift_holomorphic}):
\bq
 {u_i}'(-) = u_i(-) - z | l_j + \r,
 & &
 {\bar{u}_i}'(+) = \bar{u}_i(+),
 \nonumber \\
 {u_j}'(-) = u_j(-),
 & &
 {\bar{u}_j}'(+) = \bar{u}_j(+) + z \l l_i + |.
\eq
We give a proof by induction in the number of external particles.

We first show that the three-point functions vanish for $z \rightarrow \infty$.
We start with the pure gluon case.
$A_3({g_i'}^+,{g_j'}^+,g_k^+)$ vanishes identically, whereas $A_3({g_i'}^+,{g_j'}^+,g_k^-)$ as a function of $z$
is given by
\bq
 A_3({g_i'}^+,{g_j'}^+,g_k^-) & = & i \sqrt{2} \frac{[ji]^3}{[ki] \left( [jk] + z [ik] \right)}.
\eq
Clearly, this function vanishes for $z \rightarrow \infty$.
\\
\\
Let us now consider the case, where particle $i$ is a gluon and particle $j$ is a quark.
Then the third particle $k$ is necessarily an anti-quark. For a massive fermion line we have to consider both helicities
for particle $k$. A short calculation shows that with the choice of reference spinors 
as in eq.~(\ref{choice_of_reference_spinors}) we have
\bq
 A_3({g_i'}^+,{Q_j'}^+,\bar{Q}_k^-) 
 =
 A_3({g_i'}^+,\bar{Q}_j'{}^+,Q_k^-)
 = 0,
 \nonumber \\
 A_3({g_i'}^+,{Q_j'}^+,\bar{Q}_k^+)
 =
 A_3({g_i'}^+,\bar{Q}_j'{}^+,Q_k^+)
 = 0.
\eq
These amplitudes certainly vanish for $z \rightarrow \infty$. 
On the other it can be shown that the amplitudes
$A_3(\bar{Q}_i'{}^+,g_j'{}^+,Q_k^-)$ and
$A_3(Q_i'{}^+,g_j'{}^+,\bar{Q}_k{}^-)$ do not vanish for $z \rightarrow \infty$.
If the quark is massive, the same holds for the amplitudes
$A_3(\bar{Q}_i'{}^+,g_j'{}^+,Q_k^+)$ and 
$A_3(Q_i'{}^+,g_j'{}^+,\bar{Q}_k{}^+)$.
This places the constraint that if particle $i$ is a quark or an anti-quark, particle $j$ is a gluon
and the two are adjacent, then the helicity combination $(i^+,j^+)$ cannot be calculated with
the holomorphic shift.
\\
\\
There are a few 4- and 5-point amplitudes, which we treat separately:
\bq
\label{exceptions}
 A_4(g_i^+,g_j^+,Q, Q),
 \;\;\;
 A_4(Q_i^+,g_j^+,Q, g),
 \;\;\;
 A_5(Q_i^+,g_j^+,Q, Q', Q'),
 \;\;\;
 A_4(Q_i^+, Q_j'{}^+,Q, Q').
\eq
Here $Q$ and $Q'$ stands either for a quark or an anti-quark and no particular cyclic order is implied.
The amplitudes in eq.~(\ref{exceptions}) are the only four- or higher-point
amplitudes, where we cannot choose in addition to the marked particles
$i$ and $j$ two additional particles $k$ and $l$ such that
in the set
\bq
\label{condition_fermion_line}
 \{ i, k, l \}
\eq
no fermion line connects two of the three external particles.
These cases are discussed in appendix~\ref{sect:exceptionalcases}
and give rise to the following constraints:
The holomorphic shift cannot be used for the combination $(Q_i^+,g_j^+)$.
For the combination $(Q_i^+,Q_j'{}^+)$ the holomorphic shift can only be used if $m_i=0$.
\\
\\
We now proceed by induction in the number of external particles.
We can assume that there are two additional particles $k$ and $l$.
Since we excluded the special cases in eq.~(\ref{exceptions}), we can also assume that
in the set $\{i,k,l\}$ no two particles belong to the same fermion line.
We first discuss the case, where we can choose the two additional particles with
identical helicities.
These are the sub-cases
\bq
(i^+,j^+,k^-,l^-), \;\;\; \mbox{and} \;\;\;
(i^+,j^+,k^+,l^+).
\eq
In these cases we first consider a supplementary recursion relation, which will provide us with
an expression of the amplitude from which we can deduce the large $z$ behaviour.
This leaves the sub-case, where we cannot choose two additional particles with equal helicity assignments.
In this case $k$ and $l$ have opposite helicities and 
after a possible relabelling $k \leftrightarrow l$ we can assume that the helicity assignment is
$(i^+,j^+,k^+,l^-)$. 
We discuss this sub-case separately.
\\
\\
We first consider the case $(i^+,j^+,k^-,l^-)$.
As above we fix as reference spinors $|q_i +\r = |l_j +\r$
and $\l q_j +| = \l l_i +|$, while $\l q_i +|$ and $|q_j +\r$ are arbitrary.
For particles $k$ and $l$ we choose as reference spinors
\bq 
 | q_k+ \r = | q_l+ \r = |l_j +\r,
 & &
 \l q_k+ | = \l q_l+ | = \l l_i +|.
\eq
This choice defines $p_k^\flat$ and $p_l^\flat$.
We now consider the shift
\bq
 u_i'(-) & = & u_i(-) - z | l_j + \r - y \beta_k | p^\flat_k+ \r - y \beta_l | p^\flat_l+ \r,
 \nonumber \\
 \bar{u}_j'(+) & = & \bar{u}_j(+) + z \l l_i+ |,
 \nonumber \\
 \bar{u}_k'(+) & = & \bar{u}_k(+) + y \beta_k \l l_i+ |,
 \nonumber \\
 \bar{u}_l'(+) & = & \bar{u}_l(+) + y \beta_l \l l_i+ |,
\eq
where
$\beta_k$ and $\beta_l$ are chosen as
\bq
 \beta_k = \frac{\l p_l^\flat l_j \r}{\l p_l^\flat p_k^\flat \r},
 & &
 \beta_l = \frac{\l p_k^\flat l_j \r}{\l p_k^\flat p_l^\flat \r}.
\eq
The coefficients are chosen such that
\bq
 \beta_k | p^\flat_k+ \r + \beta_l | p^\flat_l+ \r
 & = & | l_j + \r.
\eq
The function $A(y)$ vanishes at infinity, as each individual diagram vanishes at infinity.
The argument is similar to the one we gave above for the helicity combination $(i^+,j^-)$:
In any diagram the $y$-dependence flows through a three-legged path with end-point $i$, $k$ and $l$.
Suppose first that all three particles are gluons.
The most dangerous diagrams are the ones, where we have only three-gluon-vertices along the path.
Then the combination of propagators and vertices gives a factor $y$ in the large $y$-behaviour, while
the polarisation vectors contribute a factor $1/y^3$. 
In total this diagram goes like $1/y^2$ in the large $y$ limit and therefore vanishes
as $y$ goes to infinity.
If we replace internally a gluon propagator by a quark propagator, we have to change at least
two three-gluon vertices into quark-gluon vertices.
This improves the estimate by a factor $1/y$.
If an external gluon is replaced by a fermion, we have to change at least one three-gluon vertex into a
quark-gluon vertex. This does not modify the large $y$ behaviour.
Note that we have excluded the case, where a fermion line connects two of the three particles $i$, $k$ and $l$.
\\
\\
From the fact that $A(y)$ vanishes for $y \rightarrow \infty$ we obtain 
the recursion relation
\bq
\label{recursion_y}
 A(y=0,z)
 & = & 
 \sum\limits_{\alpha,\lambda}
 A_L(y_\alpha,z,\lambda) \frac{i}{P_\alpha(z)^2-m_\alpha^2} A_R(y_\alpha,z,-\lambda),
\eq
where we dropped arguments not relevant to the discussion here.
We will use this formula to estimate the $z$-behaviour at infinity.
Suppose that $i$ and $j$ are on opposite sides of the propagator.
Then 
\bq
 P_\alpha(z)^2 & = & P_\alpha^2-z \l l_i+ | P\!\!\!/_\alpha | l_j+ \r
\eq
and
$y_\alpha$ depends linearly on $z$. If both $k$ and $l$ are on the same side as particle $j$ we have
\bq
 y_\alpha & = &
  \frac{P_\alpha^2-m_\alpha^2-z \l l_i+ | P\!\!\!/_\alpha | l_j+ \r}
       {\beta_k \l l_i+ | P\!\!\!/_\alpha | p_k^\flat + \r + \beta_l \l l_i+ | P\!\!\!/_\alpha | p_l^\flat + \r}
 =
  \frac{P_\alpha^2-m_\alpha^2}{\l l_i+ | P\!\!\!/_\alpha | l_j+ \r} - z.
\eq
A similar formula holds if only one of the particles $k$ or $l$ is on the same side as particle $j$.
The $z$-dependence flows through a four-legged path
and one can show by a diagrammatic analysis that each diagram vanishes for $z \rightarrow \infty$.
We first observe that the product of the scalar propagator, on which the amplitude is factorised, times
the two polarisation vectors attached to it, behaves like an internal propagator in the large $z$ limit. 
Let us again start from the pure gluon case and assume the worst-case scenario of only three-gluon
vertices. The product of propagators and vertices gives a factor $z$, the
three polarisation vectors for particles $i$, $k$ and $l$ contribute a factor $1/z^3$, the polarisation
vector for particle $j$ a factor $z$. In total the amplitude behaves like $1/z$ and vanishes in the 
large $z$ limit.
Replacing an internal gluon propagator by a quark propagator improves the estimate by a factor $1/z$.
Replacing an external gluon by a quark does not change the large $z$-behaviour, as long as
we do not have a fermion line connecting two of the four external particles $i$, $j$, $k$ and $l$.
As above, the cases where a fermion line connects two of the three particles $i$, $k$ and $l$
are excluded. In addition we have excluded from the very beginning the case where a fermion line
connects $i$ and $j$.
Therefore the only possibilities, where a fermion line connects two particles are the ones where
a fermion line connects particle $j$ either with particle $k$ or $l$.
In this case the total contribution from this fermion line behaves like $z$ for 
$z \rightarrow \infty$, while the rest of the diagram gives at least a factor $1/z^2$.
\\
\\
Suppose now that particles $i$ and $j$ are on the same side of the propagator, say they are both in $A_L$.
Then $y_\alpha$ is independent of $z$.
The reference spinors for particle $i$ are given by
$| l_j+\r$ and an arbitrary $\l q_i+ |$.
For particle $j$ the reference spinors are $|q_j + \r$ and $\l l_i+ |$.
Since $A_L$ has fewer legs than $A$ we can use the induction hypothesis and therefore $A_L$ vanishes
as $z$ goes to infinity.
This completes the proof for the case $(i^+,j^+,k^-,l^-)$.
\\
\\
We now consider the case $(i^+,j^+,k^+,l^+)$.
As reference spinors for particles $k$ and $l$ we take as above
\bq 
 | q_k+ \r = | q_l+ \r = |l_j +\r,
 & &
 \l q_k+ | = \l q_l+ | = \l l_i +|.
\eq
We consider the shift
\bq
 u_i'(-) & = & u_i(-) - z | l_j + \r + y [ p^\flat_k p^\flat_l ] | l_j + \r,
 \nonumber \\
 \bar{u}_j'(+) & = & \bar{u}_j(+) + z \l l_i+ |,
 \nonumber \\
 u_k'(-) & = & u_k(-) + y [ p^\flat_l l_i ] | l_j + \r,
 \nonumber \\
 u_l'(-) & = & u_l(-) + y [ l_i p^\flat_k ] | l_j + \r.
\eq
Momentum conservation is satisfied due to the Schouten
identity.
The shift in $y$ is chosen such that each individual diagram vanishes for $y \rightarrow \infty$.
Again we can show with the same steps as in the $(i^+,j^+,k^-,l^-)$ case that $A(z)$ vanishes for
$z \rightarrow \infty$.
\\
\\
Finally, we discuss the case $(i^+,j^+,k^-,l^+)$.
Assume first that particles $i$ and $j$ are massless particles. 
Then the amplitude is independent of the choice of the reference spinors for particles $i$ and $j$.
As reference spinors for particles $k$ and $l$ we take again
\bq 
 | q_k+ \r = | q_l+ \r = |l_j +\r,
 & &
 \l q_k+ | = \l q_l+ | = \l l_i +|.
\eq
We consider the shift
\bq
\label{plus_minus_shift}
 u_i'(-) & = & u_i(-) - z | l_j + \r - y | p^\flat_k+ \r,
 \nonumber \\
 \bar{u}_j'(+) & = & \bar{u}_j(+) + z \l l_i+ | - y \l p^\flat_l+ |,
 \nonumber \\
 \bar{u}_k'(+) & = & \bar{u}_k(+) + y \l l_i+ |,
 \nonumber \\
 u_l'(-) & = & u_l(-) + y | l_j+ \r.
\eq
We can show with the same steps as in the $(i^+,j^+,k^-,l^-)$ case that $A(z)$ vanishes for
$z \rightarrow \infty$.
Note that for particle $i$ and $j$ the shift in $y$ is not proportional to the reference spinors
of these particles.
Therefore the shift in eq.~(\ref{plus_minus_shift}) is restricted to massless particles.
This leaves the cases, where particle $i$ or particle $j$ or both are massive particles.
In accordance with eq.~(\ref{condition_fermion_line}) particles $k$ and $l$ are chosen such that
in the set $\{i,k,l\}$ no fermion line connects two of the three external particles.
There are only very few cases where $k$ and $l$ must be chosen such that they have opposite helicities.
These are the cases
\bq
\label{exceptions_2}
 A_4(g_i^+,Q_j^+,Q^\pm, g^\mp),
\;\;\;
 A_5(q_i^+,Q'_j{}^+,q^-,Q'{}^\pm,g^\mp),
\;\;\;
 A_6(q_i^+,Q'_j{}^+,q^-,Q'{}^\pm,Q''{}^\mp,Q''{}^\mp).
\eq
Here $q_i^+$ denotes a massless quark, since the combination $(Q_i^+,Q'_j{}^+)$ 
where $Q_i^+$ is a massive quark
is already excluded.
All cases are discussed explicitly in appendix~\ref{sect:exceptionalcases}.
It will turn out that these cases do not lead to additional restriction on the validity
of the recursion relation.

\subsection{Amplitudes involving only massive quarks or anti-quarks}
\label{sect:only_massive_quarks}

The holomorphic and anti-holomorphic two-particle shifts in eq.~(\ref{shift_holomorphic})
and eq.~(\ref{shift_antiholomorphic}) allow us to calculate recursively all amplitudes
except the ones, which consist solely of massive quarks or anti-quarks.
Among those, the four-parton amplitudes $A_4(\bar{Q},Q,\bar{Q}',Q')$ are given by just one
Feynman diagram and therefore are most efficiently calculated by a Feynman diagram
calculation.
Also the six-quark amplitudes are relatively simple. 

We consider now the ones with more than six particles.
We select two particles $i$ and $j$, not belonging to the same fermion line.
As reference spinors for particle $i$ we choose
\bq
 | q_i+ \r = | l_j + \r,
 & &
 \l q_i+ | = \l l_j + |,
\eq
while for particle $j$ we choose
\bq
 | q_j+ \r = | l_i + \r,
 & &
 \l q_j+ | = \l l_i + |.
\eq
For all other particles we choose as reference spinors
\bq
 | q_k+ \r = | l_j + \r,
 & &
 \l q_k+ | = \l l_i + |.
\eq
The helicity combination $(i^+,j^-)$ can be calculated with the holomorphic shift
eq.~(\ref{shift_holomorphic}),
while the combination $(i^-,j^+)$ can be calculated with the anti-holomorphic shift
eq.~(\ref{shift_antiholomorphic}).
This leaves the combinations $(i^+,j^+)$ and $(i^-,j^-)$. We consider first the combination
$(i^+,j^+)$.
As we are considering amplitudes with at least eight external particles, we can always find two particles
$k$ and $l$, such that in the set $\{i,k,l\}$ no fermion line connects two of the three particles and that
in addition the particles $k$ and $l$ have the same helicity assignment.
For the helicity combination $(i^+,j^+,k^-,l^-)$ we can use the shift
\bq
 u_i'(-) & = & u_i(-) - z \beta_k | p^\flat_k+ \r - z \beta_l | p^\flat_l+ \r,
 \nonumber \\
 \bar{u}_k'(+) & = & \bar{u}_k(+) + z \beta_k \l l_i+ |,
 \nonumber \\
 \bar{u}_l'(+) & = & \bar{u}_l(+) + z \beta_l \l l_i+ |,
\eq
with
\bq
 \beta_k = \frac{\l p_l^\flat l_j \r}{\l p_l^\flat p_k^\flat \r},
 & &
 \beta_l = \frac{\l p_k^\flat l_j \r}{\l p_k^\flat p_l^\flat \r}.
\eq
This is just the three-particle shift we used to establish the supplementary recursion relation in section~\ref{supplementary_recursion}.
For the helicity combination $(i^+,j^+,k^+,l^+)$ we can use the shift
\bq
 u_i'(-) & = & u_i(-) + z [ p^\flat_k p^\flat_l ] | l_j + \r,
 \nonumber \\
 u_k'(-) & = & u_k(-) + z [ p^\flat_l l_i ] | l_j + \r,
 \nonumber \\
 u_l'(-) & = & u_l(-) + z [ l_i p^\flat_k ] | l_j + \r.
\eq
Similar considerations apply to the helicity combination $(i^-,j^-)$.

\section{Applications}
\label{sect:applications}

In this section we present a few examples and applications. 
We discuss helicity amplitudes with a pair of massive quarks,
zero or one negative
helicity gluons
and an arbitrary number of positive helicity gluons.
Helicity amplitudes with a pair of massive quarks plus three gluons can be found in~\cite{Bernreuther:2004jv}.

\subsection{Amplitudes with positive helicity gluons}
In this section we consider amplitudes with one massive quark pair and
an arbitrary number of positive helicity gluons.  These amplitudes
are the building blocks for the construction of amplitudes with
negative gluons using on-shell recursion relations.  While the
amplitudes for a pair of massive scalars or quarks and an arbitrary
number of positive helicity gluons are known in closed
form~\cite{Forde:2005ue,Rodrigo:2005eu,Schwinn:2006ca,Ferrario:2006np}, 
they serve as a first example to demonstrate the application of
the shifts of momenta of massive quarks. 
Previous calculations of such amplitudes considered the shift of two gluons.

If the same spin axis is chosen for the two quarks, there are three
non-vanishing amplitudes~\cite{Schwinn:2005pi}: the helicity conserving
amplitudes $A_n(Q_1^\pm, g_2^+,\dots, \bar Q_n^\mp)$, and a helicity flip
amplitude $A_n(Q_1^-, g_2^+,\dots, \bar Q_n^-)$.
The amplitude $A_n(Q_1^-, g_2^+,\dots, \bar Q_n^+)$ is related by charge conjugation to
the amplitude $A_n(Q_1^+, g_2^+,\dots, \bar Q_n^-)$.
As discussed in section~\ref{sect:proof} both the helicity conserving 
and the helicity flip amplitudes can be computed
 applying the holomorphic shift~\eqref{shift_holomorphic} with $(i,j)=(2,1)$. 
This implies that $p_2$ is chosen as reference momentum for $Q_1$ and $\bar Q_n$,
but using the transformation~\eqref{rotate_spin_1} it is straightforward
to obtain the results for an arbitrary polarisation.
The recursion relation consists of a single term
 \begin{equation}
\label{eq:++recursion}
         A_n(Q_1^\pm,g_2^+,\dots, \bar Q_n^-)=
  A_{n-1}({{Q'}_{1}}^\pm,{g'_{23}}^{+},g_4^+,\dots, \bar Q_n^-)         
       \frac{i}{{p_{2,3}}^2}A_3( {g'_{(-23)}}^-,
       {g'_{2}}^+,g_{3}^+),
 \end{equation}
with $p_{2,3}=p_2+p_3$, 
since the degree zero amplitudes with more than three gluons vanish
on-shell.

The amplitudes with a massive quark pair with
the same spin quantisation axis are related through super-symmetric Ward identities to
amplitudes
of massive
scalars~\cite{Schwinn:2006ca}:
\begin{align}
  A_n(Q_1^+,g_2^+,\dots,\bar Q_n^-)&=\frac{\braket{p_n^\flat q}}{\braket{p_1^\flat q}}
    A_n(\phi_1^+,g_2^+,\dots,\bar \phi_n^-),
\label{eq:conserve+++}\\
A_n(Q_1^-,g_2^+,\dots,\bar Q_n^-)&=
\frac{\braket{p_1^\flat p_n^\flat}}{m} A_n(\phi_1^+,g_2^+\dots,\bar \phi_n^-).
\label{eq:flip+++}
\end{align}
The scalar amplitudes satisfy therefore the recursion relation
 \begin{equation}
\label{eq:scalar++recursion}
         A_n(\phi_1^+,g_2^+,\dots, \bar \phi_n^-)=
  A_{n-1}({{\phi'}_{1}}^+,{g'_{23}}^{+},g_4^+,\dots, \bar \phi_n^-)         
       \frac{i}{{p_{2,3}}^2}A_3( {g'_{(-23)}}^-,
       {g'_{2}}^+,g_{3}^+).
 \end{equation}
The light-like momenta $p_1^\flat$ and $p_n^\flat$ associated to $p_1$ and $p_n$ are given
by
\bq
 p_1^\flat = p_1 - \frac{m^2}{2p_1p_2} p_2,
 & &
 p_n^\flat = p_n - \frac{m^2}{2p_2p_n} p_2.
\eq
The spinors are shifted as
\bq
 |2+ \r \rightarrow |2+ \r - z | p_1^\flat + \r,
 & &
 \bar{u}_1(+) \rightarrow \bar{u}_1(+) + z \l 2 + |,
\eq
where
\bq
 z & = & \frac{p_{2,3}^2}{\l 2+ | \fmslash p_3 | p_1^\flat + \r}
 = \frac{\l 3 2 \r}{\l 3 p_1^\flat \r}.
\eq
Expressions containing the intermediate shifted momentum $p'_{2,3}$  can be 
simplified similar to the massless case~\cite{Britto:2004ap}
\bq
\label{eq:21-simplify}
  \ket{{p'_{2,3}}^\flat-}=\ket{p^\flat_{2,3}-}=\frac{\fmslash p_{2,3}\ket{p_1^\flat +}}{\braket{p^\flat_{2,3}p_1^\flat}}
  =\frac{\fmslash p_{2,3}\fmslash p_1\ket{2-}}{
\braket{p^\flat_{2,3}-|\fmslash p_1|2-}},
  &&
  \ket{{p'_{2,3}}^\flat+}=\ket{p^\flat_{2,3}+}=\frac{\fmslash p_{2,3}\ket{2-}}{\sbraket{p^\flat_{2,3}2}}.
\;\;\;\;\;\;
\eq
A particular
compact form of the scalar
amplitudes has been obtained in~\cite{Ferrario:2006np}:
\begin{equation}
\begin{aligned}
\label{eq:rodrigo}
  A_n(\phi_1^+,g_2^+,\dots,\bar \phi_n^-)
    &= 2^{n/2-1} i m^2 \frac{  \braket{2+|\prod_{j=3}^{n-2}
\left(y_{1,j}-\fmslash p_j\fmslash p_{1,j-1}\right)|(n-1)-} }{
y_{1,2} y_{1,3}\dots y_{1,n-2} 
\braket{23}\braket{34}\dots\braket{(n-2)(n-1)}},
\end{aligned}
\end{equation}
where 
\bq p_{1,j}=\sum_{1}^{j}p_j,& &y_{1,j}=p_{1,j}^2-m^2. \eq 
It is an instructive exercise to verify that eq.~(\ref{eq:rodrigo})
is a solution of eq.~(\ref{eq:scalar++recursion}):
\begin{multline}
  A_n(\phi_1^+,g_2^+,\dots, \bar \phi_n^-)
=   A_{n-1}({{\phi'}_{1}}^+,{g'_{23}}^{+},g_4^+,\dots, \bar \phi_n^-)         
       \frac{i}{{p_{2,3}}^2}A_3( {g'_{(-23)}}^-,
       {g'_{2}}^+,g_{3}^+)\\
    = 2^{n/2-1} i^3 m^2 
\frac{\braket{p^\flat_{2,3}+|\prod_{j=4}^{n-2}
\left(y_{1,j}-\fmslash p_j\fmslash p_{1,j-1}\right)|(n-1)-} }{
 y_{1,3}\dots y_{1,n-2} 
  \braket{p^\flat_{2,3}4}\braket{45}\dots\braket{(n-2)(n-1)}} 
\frac{1}{\braket{23}\sbraket{32}}
 \frac{\sbraket{32}^3}{
\sbraket{-p^\flat_{2,3} 3}\sbraket{2 (-p^\flat_{2,3})}} \\
   = 2^{n/2-1} i m^2 
\frac{\braket{2+|\fmslash p_1\fmslash p_{2,3}\prod_{j=4}^{n-2}
\left(y_{1,j}-\fmslash p_j\fmslash p_{1,j-1}\right)|(n-1)-} \sbraket{32}^2 }{
 y_{1,3}\dots y_{1,n-2} 
  \braket{23} \braket{2+|\fmslash p_{2,3}|4+}
\braket{45}\dots\braket{(n-2)(n-1)}
\braket{2+|\fmslash p_1\fmslash p_{2,3}| 3-}}\\
= 2^{n/2-1} i m^2 
\frac{\braket{2+|\prod_{j=3}^{n-2}
\left(y_{1,j}-\fmslash p_j\fmslash p_{1,j-1}\right)|(n-1)-}  }{
 y_{1,2}\dots y_{1,n-2} 
  \braket{23}\dots\braket{(n-2)(n-1)}}.
\end{multline} 
In the last step we have used the identity~\cite{Ferrario:2006np}
\begin{equation}
\label{eq:rodrigo-simplify}
\bra{2+}\fmslash p_1\fmslash p_{2,3}= 
\bra{2+}(y_{1,3}-\fmslash p_{3}\fmslash p_{1,2})
\end{equation}
to extend the product in the numerator down to $j=3$.
This example shows that the shift of a massive
quark leads to a computation similar to one for 
massless particles.


\subsection{Amplitudes with one negative helicity gluon 
adjacent to a massive quark}
\label{sec:Q-++} 

In this section we consider  amplitudes 
\bq
 A_n(Q_1^{\lambda_1},g_2^-,g_3^+,\dots,g_{n-1}^+,\bar Q_n^{\lambda_n})
\eq
with a pair of massive quarks,
a gluon with  negative helicity adjacent to a quark and  
an arbitrary number of positive helicity gluons.
As reference spinors for the massive quarks we choose
\bq
 | q_1+ \r = | q_n+ \r = | 2+ \r,
 & &
 \l q_1+ | = \l q_n + | = \l 2+ |.
\eq
The light-like momenta $p_1^\flat$ and $p_n^\flat$ associated to $p_1$ and $p_n$ are given
by
\bq
 p_1^\flat = p_1 - \frac{m^2}{2p_1p_2} p_2,
 & &
 p_n^\flat = p_n - \frac{m^2}{2p_2p_n} p_2.
\eq
For the recursion relation we consider the holomorphic shift~\eqref{eq:holomorphic} with
$(i,j)=(1,2)$.
The spinors are shifted as
\bq
\label{eq:12-shift}
 u_1(-) \rightarrow u_1(-) - z |2+ \r,
 & &
 \l 2+ | \rightarrow \l 2+ | + z \l p_1^\flat + |.
\eq
The recursion relation reads
\bq
\label{eq:bcfw-12}
\lefteqn{
 A_n(Q_1^{\lambda_1},g_2^-,g_3^+,\dots,g_{n-1}^+,\bar Q_n^{\lambda_n}) = } & & \nonumber \\
 & &  
  \sum_{j=3}^{n-1}
   A_{n-j+2}({Q'}_{1}^{\lambda_1},{g_{2,j}'}^+,g_{j+1}^+,\dots,g_{n-1}^+,\bar Q_n^{\lambda_n})
         \frac{i}{p_{2,j}^2}
   A_j({g_{-(2,j)}'}^-, {g'_{2}}^-,g_3^+,\dots,g_{j}^+)
\eq
where in the $j$'th term
\begin{equation}
  z_j=-\frac{p_{2,j}^2}{\braket{p_1^\flat + |\fmslash p_{2,j}|2+ }}.
\end{equation}
The only ingredients entering the recursion relation~\eqref{eq:bcfw-12} 
are the gluonic
MHV amplitudes and the quark amplitudes with 
positive helicity gluons~\eqref{eq:rodrigo}.
The unknown functions do not enter themselves on the right hand side, 
in contrast to the relations obtained from shifts of gluon momenta~\cite{Forde:2005ue}.
In writing~\eqref{eq:bcfw-12}
 we have used that the degree zero  gluon amplitudes with more than 
three external legs vanish on-shell and that the three point degree
zero  vertex vanishes if an anti-holomorphic spinor is shifted.

From super-symmetric Ward identities we obtain \cite{Schwinn:2006ca}
\begin{align}
\label{eq:neg-swi1}
 A(Q_1^+,g_2^+,\dots,g_j^-,\dots, \bar Q_n^+) &=0,\\
 A( Q_1^+,g_2^+,\dots,g_j^-,\dots, \bar Q_n^-) &=
\frac{\braket{p_n^\flat j}}{\braket{p_1^\flat j}}  
A_n(\phi_1^+,g_2^+,\dots,g_j^-,\dots, \bar \phi_n^-),
\label{eq:neg-swi2}\\
 A(Q_1^-,g_2^+,\dots,g_j^-,\dots, \bar Q_n^+) &=
-\frac{\braket{p_1^\flat j}}{\braket{p_n^\flat j}}  
A_n(\phi_1^-,g_2^+,\dots,g_j^-,\dots, \bar \phi_n^+).
\label{eq:neg-swi3}
\end{align}
Therefore the amplitude for the quark helicities $(Q_1^+, \bar Q_n^+)$ vanishes.
This follows also from the recursion relation~\eqref{eq:bcfw-12}. In this case the right-hand-side
of eq.~~\eqref{eq:bcfw-12} equals zero, since the
quark-gluon amplitude with only positive helicity labels vanishes.

Furthermore, eq.~(\ref{eq:neg-swi2}) and eq.~(\ref{eq:neg-swi3}) can be used to relate
the helicity combinations $(Q_1^+, \bar Q_n^-)$ and $(Q_1^-, \bar Q_n^+)$.
It follows that only the helicity combinations $(Q_1^\pm, \bar Q_n^-)$
need to be considered.

Inserting the explicit expressions for the sub-amplitudes
into~\eqref{eq:bcfw-12}
 we obtain for the helicity conserving amplitude
\begin{multline}
\label{eq:-++}
 A_n(Q_1^+,g_2^-,g_3^+,\dots, g_{n-1}^+, \bar Q_n^-)=
 2^{n/2-1} i \frac{\l p_n^\flat 2 \r}{\l p_1^\flat 2 \r} 
 \frac{1}{\braket{2 3}\dots\braket{(n-2)(n-1)}}\times\\ 
\sum_{j=3}^{n-1} 
 \frac{\braket{2-| \fmslash p_1\fmslash p_{2,j}|2+}^2}{
p_{2,j}^2 \braket{2-|\fmslash p_1\fmslash p_{2,j}|j+ }}
\left(\delta_{j,n-1}+\delta_{j\neq n-1}\frac{m^2
 \braket{2-|\fmslash p_{2,j}|\Phi_{j+1,n}-}\braket{j(j+1)}}{y_{1,j}
          \braket{2-|\fmslash p_1 \fmslash p_{2,j}|(j+1)+}}\right)
\end{multline}
where $\delta_{j\neq n-1}=1-\delta_{j,n-1}$ and we used a short-hand notation
for 
the frequently occurring quantity
\begin{equation}
\label{eq:def-phi}
\ket{\Phi_{k,n}-}= \prod_{j=k}^{n-2}
\left(1-\frac{\fmslash p_j\fmslash p_{1,j}}{y_{1,j}}\right)\ket{(n-1)-}.
\end{equation}
Intermediate expressions containing spinors of the shifted momentum
$p_{2,j}'$ have been simplified according to
\bq
  \ket{{p'_{2,j}}^\flat+}=\ket{p^\flat_{2,j}+}
  =\frac{\fmslash p_{2,j}\fmslash p_1\ket{2+}}{
\braket{p^\flat_{2,j}+|\fmslash p_1|2+}},
  &&
  \ket{{p'_{2,j}}^\flat-}=\ket{p^\flat_{2,j}-}
 =\frac{\fmslash p_{2,j}\ket{2+}}{\braket{p^\flat_{2,j}2}}.
\eq
Multiplying the result~(\ref{eq:-++}) by a factor $\braket{p_1^\flat 2}/\braket{p_n^\flat 2}$ results
in a new representation of the corresponding amplitude with a pair of
massive scalars.
Compared to a previous computation of this amplitude 
in eq. (39) of~\cite{Forde:2005ue}, our result has a similar structure
but is simpler since we used the more compact
expression~\eqref{eq:rodrigo} as input. Furthermore we obtained the result 
directly from known quantities whereas in a calculation using only
shifts of gluons~\cite{Forde:2005ue} a much more
complicated procedure of iterated shifts is necessary. 

The helicity flip amplitude is obtained with only small modifications:
\begin{multline}
\label{eq:-++flip}
 A_n(Q_1^-,g_2^-,g_3^+,\dots, g_{n-1}^+, \bar Q_n^-)=
 2^{n/2-1} i \frac{\l p_1^\flat p_n^\flat \r}{m} 
 \frac{1}{\braket{2 3}\dots\braket{(n-2)(n-1)}}\times\\ 
\sum_{j=3}^{n-1} 
 \frac{\braket{2-| \fmslash p_1\fmslash p_{2,j}|2+}^2}{
p_{2,j}^2 \braket{2-|\fmslash p_1\fmslash p_{2,j}|j+ }}
 \left( 1 + p_{2,j}^2 \frac{\braket{2 p_n^\flat}}{\braket{2-|\fmslash p_{2,j} \fmslash p_1^\flat | p_n^\flat+}} \right) \\
\times
\left(\delta_{j,n-1}+\delta_{j\neq n-1}\frac{m^2
 \braket{2-|\fmslash p_{2,j}|\Phi_{j+1,n}-}\braket{j(j+1)}}{y_{1,j}
          \braket{2-|\fmslash p_1 \fmslash p_{2,j}|(j+1)+}}\right).
\end{multline}

\section{Summary and conclusions}
\label{sect:summary}

In this paper we considered on-shell recursion relations for Born QCD amplitudes.
We put particular emphasis on amplitudes with several pairs of quarks and massive quarks
and gave a detailed description on how to shift the external particles in spinor space.
For massive quarks this implies a particular choice of reference spinors, which define the spin
quantisation axis.
We found that all Born QCD amplitudes, which have at least some external particles which are not massive quarks,
can be computed by on-shell recursion relations using two-particle shifts.
Amplitudes with only massive quarks can be computed recursively from three-particle shift.
We gave a detailed proof of the validity of the recursion relation.
As an application we considered
helicity amplitudes including a pair of massive quarks, zero or one negative
helicity gluons and an arbitrary number of positive helicity gluons. 

\subsection*{Acknowledgments}
CS was supported by the DFG Sonderforschungsbereich/Transregio 9 "Computergest\"utzte Theoretische Teilchenphysik".

\appendix

\section{Spinors}
\label{appendix:spinors}

We define the light-cone coordinates as
\bq
p_+ = p_0 + p_3, \;\;\; p_- = p_0 - p_3, \;\;\; p_{\bot} = p_1 + i p_2,
                                         \;\;\; p_{\bot^\ast} = p_1 - i p_2.
\eq
In terms of the light-cone components of a light-like four-vector, the corresponding massless spinors $\l p \pm |$ and $| p \pm \r$ 
can be chosen as
\bq
\left| p+ \right\r = \frac{e^{-i \frac{\phi}{2}}}{\sqrt{\left| p_+ \right|}} \left( \begin{array}{c}
  -p_{\bot^\ast} \\ p_+ \end{array} \right),
 & &
\left| p- \right\r = \frac{e^{-i \frac{\phi}{2}}}{\sqrt{\left| p_+ \right|}} \left( \begin{array}{c}
  p_+ \\ p_\bot \end{array} \right),
 \nonumber \\
\left\l p+ \right| = \frac{e^{-i \frac{\phi}{2}}}{\sqrt{\left| p_+ \right|}} 
 \left( -p_\bot, p_+ \right),
 & &
\left\l p- \right| = \frac{e^{-i \frac{\phi}{2}}}{\sqrt{\left| p_+ \right|}} 
 \left( p_+, p_{\bot^\ast} \right),
\eq
where the phase $\phi$ is given by
\bq
p_+ & = & \left| p_+ \right| e^{i\phi}.
\eq
If $p_+$ is real and $p_+>0$ we have the following relations between a spinor corresponding to a vector $p$
and a spinor corresponding to a vector $(-p)$:
\bq
 \left| \left(-p\right) \pm \right\r & = & i \left| p \pm \right\r,
 \nonumber \\
 \left\l \left(-p\right) \pm \right| & = & i \left\l p \pm \right|.
\eq
Therefore the
spinors of massive quarks and anti-quarks are related by
$u(-k,\pm)=i v(k,\pm)$ and $\bar u(-k,\pm)=i \bar v(k,\pm)$. The
polarisation vectors of the gluons are unchanged under the reversal of
the momentum.  
Spinor products are denoted as
\bq
 \l p q \r = \l p - | q + \r = p^A q_A,
 & &
 [ q p ] = \l q + | p - \r = q_{\dot{A}} p^{\dot{A}}.
\eq 

\section{Exceptional cases}
\label{sect:exceptionalcases}

For the exceptional cases we consider as an example the helicity configuration $(i^+,j^+)$
with the holomorphic shift.
Similar considerations apply to the anti-holomorphic shift and to the configuration $(i^-,j^-)$ with both
types of shifts.
The exceptional cases have two origins:
First, for the helicity configuration $(i^+,j^+)$ with the holomorphic shift we have to consider the cases 
where
we cannot choose to additional particles $k$ and $l$, such that
in the set
\bq
 \{ i, k, l \}
\eq
no fermion line connects two of the three external particles.
These are the cases listed in eq.~(\ref{exceptions}).
Secondly, we have to consider the cases, where particle $i$ or particle $j$ is a massive quark or anti-quark and
one cannot choose two additional particles $k$ and $l$ with equal helicities.
These are the cases listed in eq.~(\ref{exceptions_2}).

The exceptional cases are all limited to amplitudes with no more than six external particles,
We discuss these amplitudes case by case.
We start with the cases related to eq.~(\ref{exceptions})
and discuss at the end the cases of eq.~(\ref{exceptions_2}).
\\
\\
a) The case $A_4(Q,g_i^+,g_j^+,\bar{Q})$: We consider the relevant helicity amplitudes for massive quarks.
We have for the unshifted amplitudes
\bq
\label{amplitudes_case_a}
 A_4(Q_1^+,g_2^+,g_3^+,\bar{Q}_4^+) & = & 
  - 2 i \frac{m \l q_1 q_4 \r}{\l q_1 p_1^\flat \r \l p_4^\flat q_4 \r} 
          \frac{[ 2 3 ]}{\l  2 3 \r} \frac{m^2}{2p_1p_2},
 \nonumber \\
 A_4(Q_1^+,g_2^+,g_3^+,\bar{Q}_4^-) & = & 
   2 i \frac{\l q_1 - | p_4 | q_4 - \r}{\l q_1 p_1^\flat \r [ p_4^\flat q_4 ]} 
          \frac{[ 2 3 ]}{\l  2 3 \r} \frac{m^2}{2p_1p_2},
 \nonumber \\
 A_4(Q_1^-,g_2^+,g_3^+,\bar{Q}_4^+) & = & 
   - 2 i \frac{\l q_1 + | p_1 | q_4 + \r}{[ q_1 p_1^\flat ] \l p_4^\flat q_4 \r} 
          \frac{[ 2 3 ]}{\l  2 3 \r} \frac{m^2}{2p_1p_2},
 \nonumber \\
 A_4(Q_1^-,g_2^+,g_3^+,\bar{Q}_4^-) & = & 
   2 i \frac{\l q_1 + | p_1 p_4 | q_4 - \r}{[ q_1 p_1^\flat ] [ p_4^\flat q_4 ]} 
          \frac{[ 2 3 ]}{\l  2 3 \r} \frac{m}{2p_1p_2}.
\eq
The massless case is included as the special case $m=0$, in which all four helicity amplitudes vanish.
For the holomorphic shift we have the substitution
\bq
 | 2+ \r & \rightarrow & |2+ \r - z |3+ \r,
 \nonumber \\
 \l 3+ | & \rightarrow & \l 3+ | + z \l 2+ |.
\eq
One observes that all non-vanishing helicity amplitudes fall off as $1/z$ for large $z$ due to the factor
$1/(2 p_1 p_2)$.
This case does not lead to any restrictions.
\\
\\
b) The case $A_4(Q_i^+,g_j^+,g,\bar{Q})$: 
This case is already excluded, as $i$ and $j$ are adjacent. 
\\
\\
c) The case $A_4(Q_i^+,g,g_j^+,\bar{Q})$:
Apart from the helicity amplitudes $A_4(Q_1^+,g_2^+,g_3^+,\bar{Q}_4^+)$ and \\
$A_4(Q_1^+,g_2^+,g_3^+,\bar{Q}_4^-)$,
which were already given in eq.~(\ref{amplitudes_case_a}), we need the following two amplitudes:
\bq
 A_4(Q_1^+,g_2^-,g_3^+,\bar{Q}_4^+) & = & 
 2 i \frac{m \l q_1 q_4 \r}{\l q_1 p_1^\flat \r \l p_4^\flat q_4 \r}
   \frac{\l 2- | \fmslash p_4 | 3- \r}{2p_1p_2}
   \left( \frac{\l q_1 2 \r \l 2 q_4 \r}{\l 2 3 \r \l q_1 q_4 \r}
          - \frac{\l 2- | \fmslash p_4 | 3- \r}{s_{23}} \right), 
 \nonumber \\
 A_4(Q_1^+,g_2^-,g_3^+,\bar{Q}_4^-) & = & 
 \frac{2 i}{\l q_1 p_1^\flat \r [ p_4^\flat q_4 ]}
   \frac{\l 2- | \fmslash p_4 | 3- \r}{2p_1p_2}
 \nonumber \\
 & & \times 
   \left( \frac{\l 2- | \fmslash p_4 | 3- \r \l q_1- | \fmslash p_4 | q_4- \r}{s_{23}}
          - \frac{\l q_1 2 \r \l 2- | \fmslash p_4 | q_4 - \r}{\l 23 \r} \right). 
\eq
For the holomorphic shift we have $|q_1+\r=|3+\r$ and the substitution
\bq
 | p_1^\flat + \r & \rightarrow & | p_1^\flat + \r - z |3+ \r,
 \nonumber \\
 \l 3+ | & \rightarrow & \l 3+ | + z \l p_1^\flat + |.
\eq
We observe that all helicity amplitudes go to a constant for $z \rightarrow \infty$.
Therefore these helicity amplitudes cannot be computed with the holomorphic shift.
As the proof for the recursion relation for the helicity combination $(i^+,j^+)$ is based on induction,
we have to exclude for the holomorphic shift all combinations, where particle $i$ is a quark or an anti-quark
and particle $j$ is a gluon.
\\
\\
d) Five-parton amplitudes:
The five-parton amplitudes of eq.~(\ref{exceptions}) are the following:
\begin{itemize}
\item The cases $A_5(\bar{Q}_i^+,Q,g_j^+,\bar{Q}',Q')$ and $A_5(\bar{Q},Q_i^+,g_j^+,\bar{Q}',Q')$;
\item The cases $A_{5,sl}(\bar{Q}_i^+,Q,g_j^+;\bar{Q}',Q')$ and $A_{5,sl}(\bar{Q},Q_i^+,g_j^+;\bar{Q}',Q')$: 
These are partial amplitudes, where the particles
$(\bar{Q},Q,g)$ form one colour cluster, while the particles $(\bar{Q}',Q)$ form a second colour cluster.
\item The cases $A_{5,sl}(\bar{Q}_i^+,Q;g_j^+,\bar{Q}',Q')$ and $A_{5,sl}(\bar{Q},Q_i^+;g_j^+,\bar{Q}',Q')$: 
These are partial amplitudes, where the particles
$(\bar{Q}',Q',g)$ form one colour cluster and the particles $(\bar{Q},Q)$ form a second colour cluster.
\end{itemize}
In view of the conclusions from case c) above,
all these cases are already excluded, as particle $i$ is either a quark or an anti-quark, while particle $j$ is a gluon.
\\
\\
e) The cases $A_4(\bar{Q},Q_i^+,{\bar{Q}'}_j{}^+,Q')$ and $A_4(\bar{Q},Q_i^+,\bar{Q}',{Q'}_j{}^+)$:
We first consider $A_4(\bar{Q},Q_i^+,{\bar{Q}'}_j{}^+,Q')$. The relevant unshifted amplitudes are:
\bq
\lefteqn{
 A_4(\bar{Q}_1^-,Q_2^+,{\bar{Q}'}_3{}^+,Q_4'{}^-)
 = 
 \frac{2i}{\l q_2 p_2^\flat\r [ p_1^\flat q_1 ] [ q_4 p_4^\flat ] \l p_3^\flat q_3 \r (p_1+p_2)^2}
 } & & \nonumber \\
 & & \times
 \left(
   \l q_2- | \fmslash p_2 \fmslash p_3 | q_3+ \r \l q_4+ | \fmslash p_4 \fmslash p_1 | q_1 - \r
   - m_2^2 \l q_4+ | \fmslash p_4 | q_2+ \r \l q_1+ | \fmslash p_3 | q_3+ \r
 \right. \nonumber \\
 & & \left. 
   - m_3^2 \l q_2- | \fmslash p_2 | q_4- \r \l q_3- | \fmslash p_1 | q_1- \r
   + m_2^2 m_3^2 \l q_2 q_3 \r [ q_4 q_1 ]
 \right),
 \nonumber \\
\lefteqn{
 A_4(\bar{Q}_1^-,Q_2^+,{\bar{Q}'}_3{}^+,Q_4'{}^+)
 = 
 \frac{2i m_3}{\l q_2 p_2^\flat\r [ p_1^\flat q_1 ] \l q_4 p_4^\flat \r \l p_3^\flat q_3 \r (p_1+p_2)^2}
 } & & \nonumber \\
 & & \times
 \left(
   \l q_2- | \fmslash p_2 \fmslash p_3 | q_3+ \r \l q_4- | \fmslash p_1 | q_1 - \r
   + \l q_2- | \fmslash p_2 \fmslash p_4 | q_4+ \r \l q_3- | \fmslash p_1 | q_1 - \r
 \right. \nonumber \\
 & & \left. 
   + m_2^2 \l q_2 q_4 \r \l q_1+ | \fmslash p_3 | q_3+ \r
   + m_2^2 \l q_2 q_3 \r \l q_4- | \fmslash p_4 | q_1- \r
 \right),
 \nonumber \\
\lefteqn{
 A_4(\bar{Q}_1^+,Q_2^+,{\bar{Q}'}_3{}^+,Q_4'{}^-)
 = 
 \frac{2i m_2}{\l q_2 p_2^\flat\r \l p_1^\flat q_1 \r [ q_4 p_4^\flat ] \l p_3^\flat q_3 \r (p_1+p_2)^2}
 } & & \nonumber \\
 & & \times
 \left(
   \l q_2- | \fmslash p_4 | q_4- \r \l q_3- | \fmslash p_3 \fmslash p_1 | q_1 + \r
   - \l q_2- | \fmslash p_2 \fmslash p_3 | q_3+ \r \l q_4+ | \fmslash p_4 | q_1 + \r
 \right. \nonumber \\
 & & \left. 
   - m_3^2 \l q_2 q_3 \r \l q_4+ | \fmslash p_1 | q_1+ \r
   + m_3^2 \l q_3 q_1 \r \l q_2- | \fmslash p_2 | q_4- \r
 \right),
 \nonumber \\
\lefteqn{
 A_4(\bar{Q}_1^+,Q_2^+,{\bar{Q}'}_3{}^+,Q_4'{}^+)
 = 
 \frac{2i m_2 m_3}{\l q_2 p_2^\flat\r \l p_1^\flat q_1 \r \l q_4 p_4^\flat \r \l p_3^\flat q_3 \r (p_1+p_2)^2}
 } & & \nonumber \\
 & & \times
 \left(
 - \l q_2 q_4 \r \l q_3-| \fmslash p_3 \fmslash p_1 | q_1 + \r
 - \l q_4 q_1 \r \l q_2-| \fmslash p_2 \fmslash p_3 | q_3+ \r
 \right. \nonumber \\
 & & \left. 
  - \l q_2 q_3 \r \l q_4- | \fmslash p_4 \fmslash p_1 | q_1+ \r
 - \l q_3 q_1 \r \l q_2-| \fmslash p_2 \fmslash p_4 | q_4+ \r
 \right),
\eq
For the holomorphic shift we have
$| q_2+\r= |l_3+\r$ and
$\l q_3+|=\l l_2+|$.
As a consequence
\bq
\l p_2^\flat+| = \l l_2+ | & \mbox{and} & |p_3^\flat+ \r = |l_3+ \r.
\eq
We shift
\bq
 | p_2^\flat+ \r & \rightarrow & | p_2^\flat+ \r - z | l_3+ \r,
 \nonumber \\
 \l p_3^\flat+ | & \rightarrow & \l p_3^\flat+ | + z \l l_2+ |.
\eq
We can summarise the conditions under which the individual helicity amplitudes vanish for $z\rightarrow \infty$ as follows:
\\
\begin{tabular}{lllll}
& & & \\
$A_4(\bar{Q}_1^-,Q_2^+,{\bar{Q}'}_3{}^+,Q_4'{}^-)$: & $m_2=0$, &            & or $\l q_1+ | = \l l_2 +|$.& \\
$A_4(\bar{Q}_1^-,Q_2^+,{\bar{Q}'}_3{}^+,Q_4'{}^+)$: & $m_2=0$, & or $m_3=0$, & or $\l q_1+ | = \l l_2 +|$, & or $| q_4+ \r = | l_3 + \r$.\\
$A_4(\bar{Q}_1^+,Q_2^+,{\bar{Q}'}_3{}^+,Q_4'{}^-)$: & $m_2=0$.& & & \\
$A_4(\bar{Q}_1^+,Q_2^+,{\bar{Q}'}_3{}^+,Q_4'{}^+)$: & $m_2=0$, & or $m_3=0$, & & or $| q_4+ \r = | l_3 + \r$. \\
 & & & \\
\end{tabular} 
\\
We are interested in computing for the combination $(i^+,j^+)$ all helicity combinations with respect to the remaining
particles. Therefore the common requirement is $m_2=0$. In other words, for the combination
$(q_i^+,\bar{q}_j^+)$ the case where particle $i$ is a massive quark has to be excluded.

If we now consider the case $A_4(\bar{Q},Q_i^+,\bar{Q}',{Q'}_j{}^+)$, we find in complete analogy again the requirement
$m_2=0$. 
Therefore we also exclude the
combination $(q_i^+,q_j^+)$ where particle $i$ is a massive quark.  
\\
\\ 
f) The case $A_4(Q_j^+,g,g_i^+,\bar{Q})$: This is an additional case related to eq.~(\ref{exceptions_2}).
We are only interested in the case, where the two additional particles have opposite helicities.
These are the amplitudes
$A_4(Q_1^+,g_2^+,g_3^+,\bar{Q}_4^-)$ and
$A_4(Q_1^+,g_2^-,g_3^+,\bar{Q}_4^+)$.
One easily shows that both amplitudes vanish as $1/z^2$ for $z \rightarrow \infty$.
\\ 
\\ 
g) The case $A_4(Q_j^+,g_i^+,g,\bar{Q})$:
This is again a case related to eq.~(\ref{exceptions_2}).
We are only interested in the case, where the two additional particles have opposite helicities.
These are the amplitudes
$A_4(Q_1^+,g_2^+,g_3^+,\bar{Q}_4^-)$ and
$A_4(Q_1^+,g_2^+,g_3^-,\bar{Q}_4^+)$.
Both amplitudes vanish as $1/z$ for $z \rightarrow \infty$.
\\
\\
h) The cases
$A_5(q_i^+,Q'_j{}^+,q^-,Q'{}^\pm,g^\mp)$ and
$A_6(q_i^+,Q'_j{}^+,q^-,Q'{}^\pm,Q''{}^\mp,Q''{}^\mp)$.  These cases
are again related to eq.~(\ref{exceptions_2}).  There are several
partial amplitudes which we would have to consider.  In this case it
is simpler to discuss groups of Feynman diagrams and show that they
vanish in the limit $z\rightarrow \infty$.  We group the Feynman
diagrams contributing to $A_5(q_i^+,Q'_j{}^+,q^-,Q'{}^\pm,g^\mp)$ or
$A_6(q_i^+,Q'_j{}^+,q^-,Q'{}^\pm,Q''{}^\mp,Q''{}^\mp)$ into three
sets: Set 1 consists of all diagrams, where the $z$-dependence flows
through only one propagator.  Set 2 consists of all diagrams, where
the $z$-dependence flows through more than one propagator and which do
not contain a three-gluon vertex.  Finally, set 3 consists of all
diagrams which contain a three-gluon vertex.  

With arguments similar to the ones given in case e) and f) one shows
that the contribution from set 1 vanishes for $z \rightarrow
\infty$. To see this, note that the five and six-point diagrams in set
1 can be obtained from the four-quark amplitudes discussed previously
by setting $m_2=0$ and replacing one of the external spinors by an
off-shell quark current.

The contribution from set 2 vanishes for $z \rightarrow \infty$ since there are at least two
$z$-dependent propagators and no $z$-dependent vertices.
Finally, a short calculation reveals that also the contribution from set 3 vanishes for
$z\rightarrow \infty$.


\providecommand{\href}[2]{#2}\begingroup\raggedright\endgroup

\end{document}